\documentclass[prl,showpacs,english,amsmath,notitlepage,twocolumn]{revtex4-1}

\usepackage{color,graphicx}
\usepackage{ulem}

\begin{document}

\title{Long Range Triplet Josephson Current and $0-\pi$ Transition in Tunable Domain Walls}
\author{Thomas E. Baker, Adam Richie-Halford, and Andreas Bill}
\email[Author to whom correspondence should be addressed: ]{abill@csulb.edu}
\affiliation{Department of Physics \& Astronomy, California State University Long Beach, Long Beach,, CA 90840}

\begin{abstract}
The order parameter of superconducting pairs penetrating an inhomogeneous magnetic material can acquire a long range triplet component (LRTC) with non-zero spin projection.  This state has been predicted and generated recently in proximity systems and Josephson junctions.  We show using an analytically derived domain wall of an exchange spring how the LRTC emerges and can be tuned with the twisting of the magnetization.  We also introduce a new kind of Josephson current reversal, the triplet $0-\pi$ transition, that can be observed in one and the same system either by tuning the domain wall or by varying temperature.
\end{abstract}


\maketitle

%%%%%%%%%%%%%%%%%%%%%%%%%%%%%%%%%%%%%%%%%%%
In all known three-dimensional superconductors the condensate is composed of paired spin$-1/2$ fermions with $S_z = 0$ total spin projection along the quantization axis. A decade ago, Bergeret, Volkov, and Efetov (BVE) predicted that under certain conditions a triplet component with non-zero spin projection ($S_z = \pm 1$) of the pair amplitude might arise in a magnetic material even if the adjacent superconductor (S) has singlet pairing interactions \cite{bergeretPRB00,bergeretPRL01,bergeretPRB01}. This state is different from the Fulde-Ferrell-Larkin-Ovchinnikov (FFLO) state in which a moderate magnetic field applied to a singlet S generates an $S_z = 0$ triplet state where the Cooper pairs acquire a finite center-of-mass momentum \cite{fuldePRL70,larkinJETP65}.

One of the interesting consequences of the BVE prediction is that the superconducting condensate may penetrate into a magnetic material much deeper than expected for a pure $S_z = 0$ state. This is due to the fact that this Long Range Triplet Component (LRTC) is unaffected by the internal magnetic field. The LRTC extends over distances similar to those of singlet Cooper pairs in a normal metal.
The BVE prediction has since garnered great interest
\cite{buzdinRMP05,bergeretRMP05,fominovJETPL03,fominovPRB07,fominovJETPL10,champelPRB05,champelPRB05b,ivanovPRB06,houzetPRB07,eschrigNP08,alidoustPRB10,alidoustPRB10b,guPRB10,buzdinPRB11,eschrigPT11,wuPRB12} and recent measurements of the critical current in magnetic Josepshon junctions \cite{keizerN06,khairePRL10,anwarPRB10, robinsonS10, robinsonPRL10,khasawnehSST11,klosePRL12} and the critical temperature in proximity systems \cite{zhuPRL10,leksinPRL12} confirmed the existence of the LRTC.

\begin{figure}[h]
\includegraphics[scale=0.15]{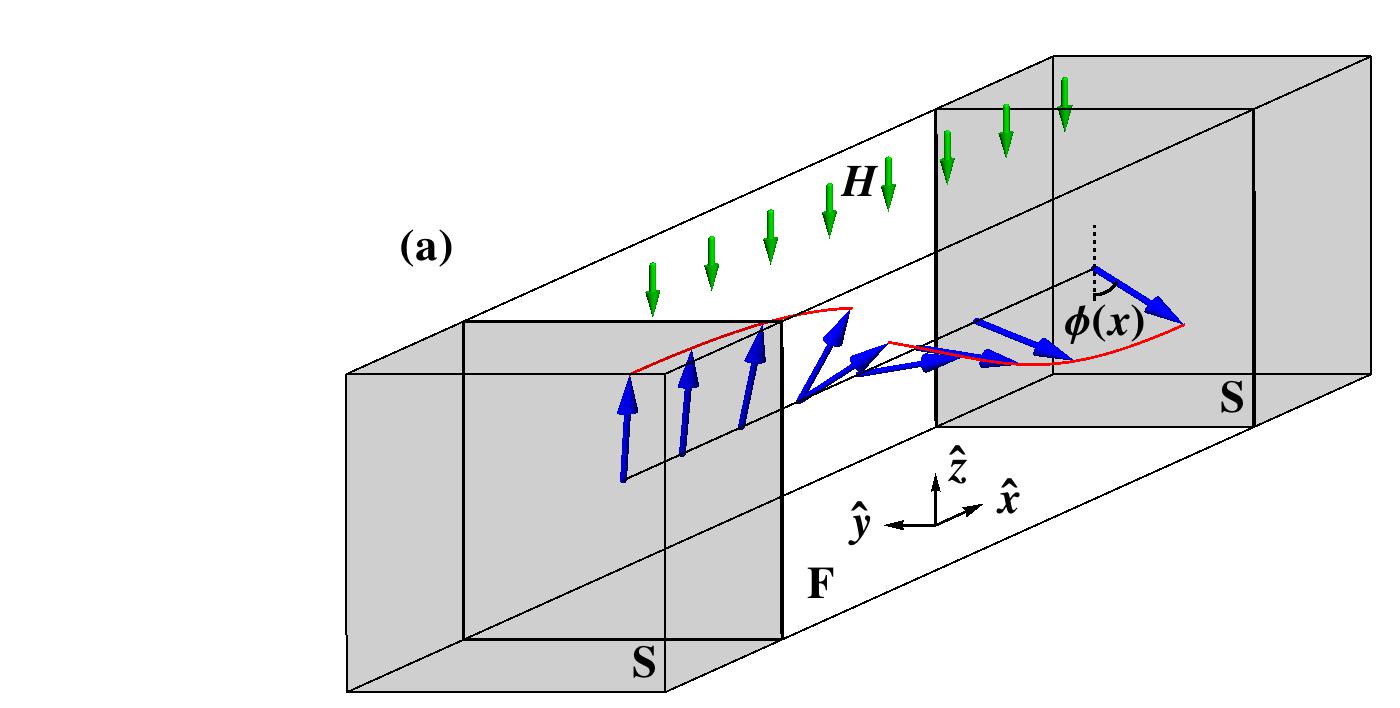}
\includegraphics[scale=0.12]{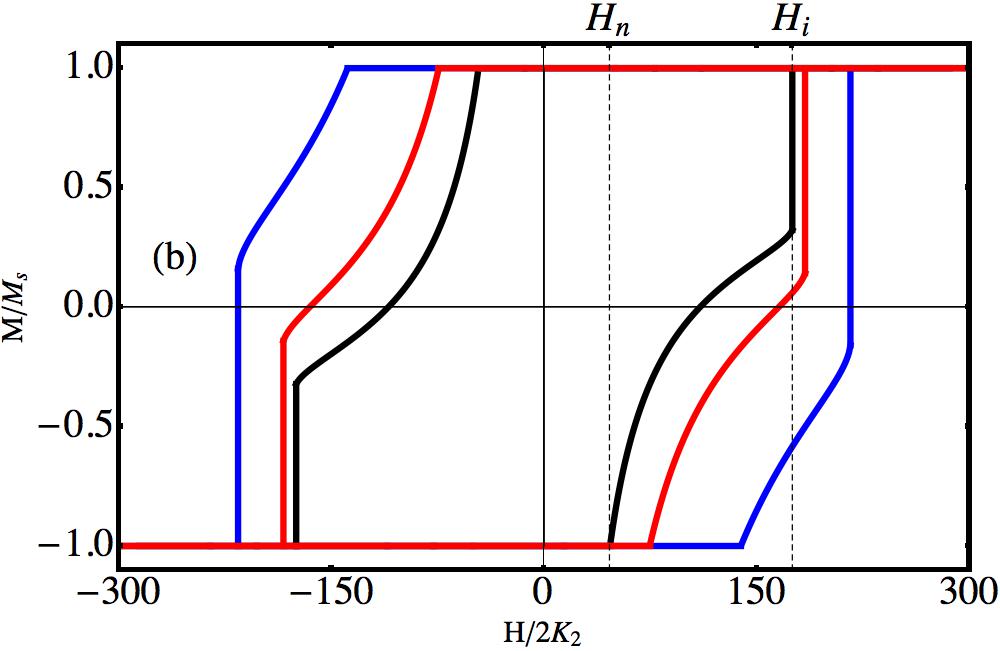}
\includegraphics[scale=0.12]{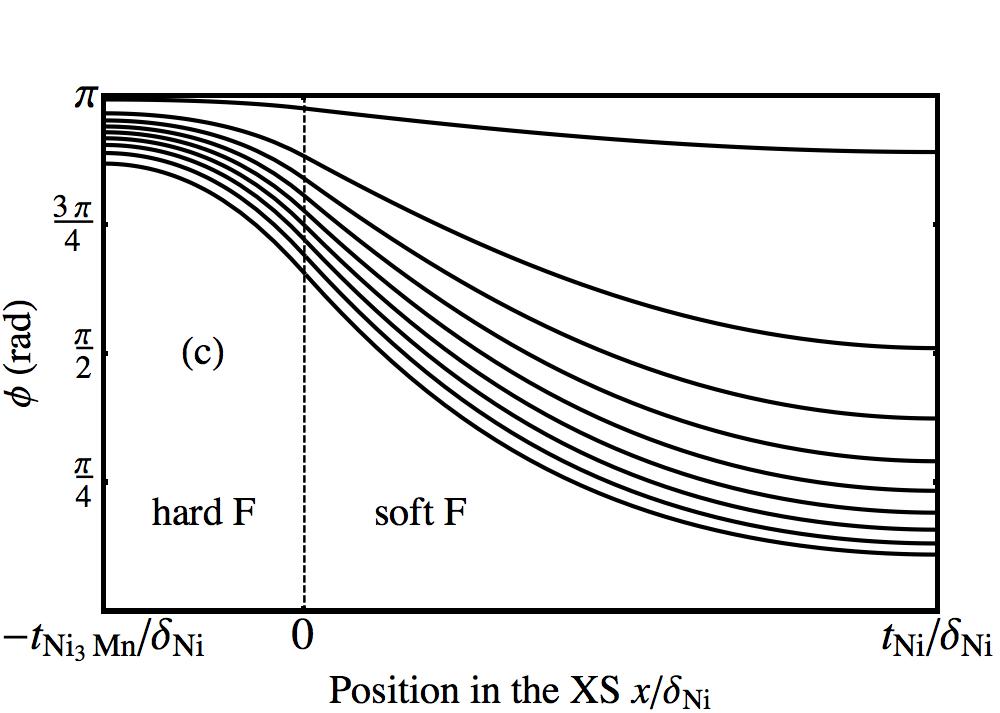}
\caption[Configuration of the ferromagnetic Josephson junction]{\label{bakerFig1} (color online). (a) Schematic ferromagnetic Josephson junction. The shaded thin films are singlet Ss. The interior is an XS composed of two Fs and the arrows display a typical domain wall (see text). (b, c) Generic magnetic properties of an XS (here Ni$_3$Mn/Ni): (b) Hysteresis loop for three thicknesses of the soft F: $t_{\rm Ni}/\delta_{\rm Ni} = 0.1$ (blue, outermost loop), 0.15 (red), 0.2 (black). The rounded part in the region $[H_n/2K_2,H_i/2K_2]$ (shown for the innermost, black loop) denotes the reversible spring region of the loop in which the domain wall exists \cite{billJMMM04}. $t_{\rm Ni_3Mn}/\delta_{\rm Ni_3Mn} = 2$. (c) Domain wall configurations for increasing values of $H/2K_2$ (from top to bottom) for the innermost hysteresis loop in (b) and represented by the angle $\phi(x)$ depicted in (a). }
\end{figure}

In this letter we use an exchange spring (XS) with an experimentally realistic and mathematically known closed form \cite{billJMMM04,bakerAJP12}, tunable magnetic domain wall to study the emergence of the LRTC of the order parameter.  We also predict three effects occurring in an S/XS/S Josephson junction. First, we show that starting with the homogeneous XS, which has no LRTC  and no Josephson current, the progressive winding up of the domain wall generates an increasing current directly attributable to the LRTC. This current is shown to appear in very wide junctions of strong ferromagnets (F) because the magnetization profile changes under the right conditions to convert short range ($S_z = 0$) to long range ($S_z = \pm 1$) components of the order parameter. Second, we propose a new kind of Josephson current reversal, the {\it triplet} $0-\pi$ transition, that occurs because the current contribution emerging from the LRTC with the tuning of the domain wall overcomes the current produced by the singlet component.  Third, we observe that for specific conditions and twist of the domain wall the {\it triplet} $0-\pi$ transition can be induced by varying temperature.
Previously both theory and experiment studied the $0-\pi$ transition of a singlet current by varying the thickness of the F over a few nanometers using the oscillatory behavior of the superconducting order parameter to reverse the direction of the singlet Josephson current \cite{buzdinJETP91,demlerPRB97,kontosPRL02,ryazanovPRL01,oboznovPRL06,pianoEPJ07}.  This requires fabricating a new sample for each thickness under identical experimental conditions. The tunable XS avoids the latter and leads to a triplet $0-\pi$ transition in wide junctions.

%%%%%%%%%%%%%%%%%%%%%%%%%%%%%%%%%%%%%%%%%%%
We consider the S/XS/S magnetic Josephson junction of Fig.~\ref{bakerFig1}a. The XS is a bilayer of homogeneous Fs with uniaxial anisotropy (along $\mathbf{\hat{z}}$) and different magnetic anisotropy energies that interact at their interface (in Fig.~\ref{bakerFig1}c we consider strong interactions between the soft and hard F and thus no phase slip at the interface). The important parameters of the XS  for the following considerations are the anisotropy constants ratio $K_1/K_2$, the intrinsic domain wall width $\delta_\alpha$ and the saturation magnetization $h_\alpha$, where $\alpha = 1, 2$ refer to the hard and soft F, respectively.
Applying a small magnetic field $H$ in direction opposite to the equilibrium magnetization induces a partial to full domain wall. In an earlier work, by one of the authors, an analytic description of the domain wall in a XS was derived \cite{billJMMM04}. That model provides an excellent description of observables such as the hysteresis loop or the magnetoresistance of experimentally realized XSs \cite{billJMMM04,mibuJMMM96,odonovanPRL02}. The great advantage of the XS is that the magnetic configuration can be smoothly tuned from a homogeneous F to a Bloch domain wall. We leverage this knowledge to offer a clear picture of the conditions under which a magnetic domain wall generates a LRTC of the superconducting order parameter.

Two XS are suggested as guidelines for which the LRTC could be measured: 1) Co/Py is a strong XS familiar in the field of magnetism that leads to small but observable triplet currents, 2) Ni$_3$Mn/Ni is proposed here as a weak XS that under certain annealing conditions presents a strong anisotropy ratio \cite{bozorth03}.  We predict such system to generate a current an order of magnitude larger than presently measured on other systems or predicted with  Co/Py.

All considerations are made in the dirty limit where the elastic scattering length is much smaller than the superconducting coherence length $\xi_S = \sqrt{\hbar D_S/2\pi T_c}$ ($D_S$ is the diffusion length of the S and $T_c$ the critical temperature of the proximity system).  In this semi-classical regime and in the absence of $S_z\neq 0$ triplet components the state of the system is determined by the Usadel equations for the scalar Green and Gor'kov  functions $g_0$ and $f_0$, respectively \cite{usadelPRL70}. In the presence of an inhomogeneous effective magnetic field, the above description has to be supplemented by vector functions $\mathbf{g}, \mathbf{f}$ to account for the possible presence of $S_z = \pm 1$  components of the order parameter. Following Ivanov and Fominov, the functions can be parametrized as \cite{ivanovPRB06}
\begin{eqnarray}\label{eq:parametrization}
\begin{cases}
g_0=M_0\cos\theta,\\
f_0=M_0\sin\theta,
\end{cases}
\quad\mathrm{and}\quad
\begin{cases}
\mathbf{g} = i \mathbf{M}\sin\theta,\\
\mathbf{f} = - i \mathbf{M}\cos\theta,
\end{cases}
\end{eqnarray}
where $M_0$ is the scalar component and $\mathbf{M} = \left( M_x,M_y,M_z \right)$ denotes the triplet amplitude vector. All five unknowns depend on position in the Ss and the F. For this parametrization the generalized Usadel equations take the form \cite{ivanovPRB06}
\begin{widetext}
\begin{eqnarray}
\label{usadeltheta} 
\frac{D}{2}\boldsymbol{\nabla}^2\theta - M_0 ( \omega_n \sin\theta - \Delta \cos\theta ) - (\mathbf{h} \cdot \mathbf{M}) \cos\theta &=& 0,\\
\label{usadelM}
\frac{D}{2}\left( \mathbf{M} \boldsymbol{\nabla}^2M_0
- M_0\boldsymbol{\nabla}^2\mathbf{M}\right)
+  \mathbf{M} \left( \omega_n \cos\theta + \Delta \sin\theta \right)
- \mathbf{h} \,M_0 \sin\theta &=& 0,
\end{eqnarray}
\end{widetext}
together with the normalization condition $M_0^2-|\mathbf{M}|^2 = 1$ \cite{zaikinZP84}.
When $\mathbf{h} = 0$, $\mathbf{M} = 0$ and $M_0 = 1$, which reduces the equations to the standard Usadel form, and when $\mathbf{h}$ is constant, then the field and $\mathbf{M}$ are aligned.
The equations are written in the Matsubara formalism with $\omega_n = (2n+1)\pi T$ ($n$ is an integer) in which case the functions are all real if $\Delta$ is real \cite{ivanovPRB06}.

The five unknown functions have to be determined self-consistently with the pair potential $\Delta(x)$ in the S and the magnetization profile in the XS
\begin{eqnarray}
\mathbf{h}_j(x) = - h_j \left[\sin\phi_j(x) \mathbf{\hat{y}} + \cos\phi_j(x) \mathbf{\hat{z}}\right]
\end{eqnarray}
where $j=1$ and 2 are the hard and soft F, respectively. The angle $\phi_j(x)$ represents the tunable domain wall of Fig.~\ref{bakerFig1}. 
The pair potential $\Delta(x)$ is defined in terms of the singlet pairing coupling constant and Gor'kov function in S. 
We consider a BCS type S in the wide dirty limit, in which case the variation of $\Delta(x)$ occurring over the coherence length, $\xi_S$, can be neglected. Consequently, $\Delta(x) \equiv \Delta_{\rm BCS}$ in the S and vanishes in the XS. 

Eqs.~(\ref{usadeltheta},\ref{usadelM}) are solved numerically with the normalization condition and usual boundary conditions at the interfaces between S and XS, and at the outer edges of the Ss \cite{zaikinZP84}. The two systems, S/XS and XS/S, are solved separately since in the wide limit the two parts are essentially decoupled and additive. 
Once the solutions of Eqs.~(\ref{usadeltheta},\ref{usadelM}) are obtained one can determine the Josephson critical current with
\begin{eqnarray}\label{IcRN}
I_c(x) = \frac{ \pi T \sigma_F}{4e} \sum_{n = -\infty}^\infty \sum_{\alpha = \{0,y,z\}}
\mathrm{Im}\left( f_{\alpha,-n}^\star\frac{\partial f_{\alpha,n}}{\partial x} \right),
\end{eqnarray}
where $\sigma_F$ is the conductivity of the ferromagnetic metal. For Bloch domain walls $f_{x,n} = 0$.

%%%%%%%%%%%%%%%%%%%%%%%%%%%%%%%%%%%%%%%%%%%
Some insight into the behavior of the LRTC can be gained from plotting the relative orientation of the Gor'kov vector function $\mathbf{f}(x)$ with respect to the magnetization $\mathbf{h}(x)$ in the XS.
\begin{figure}[htb] 
\begin{center}
\includegraphics[scale=0.15]{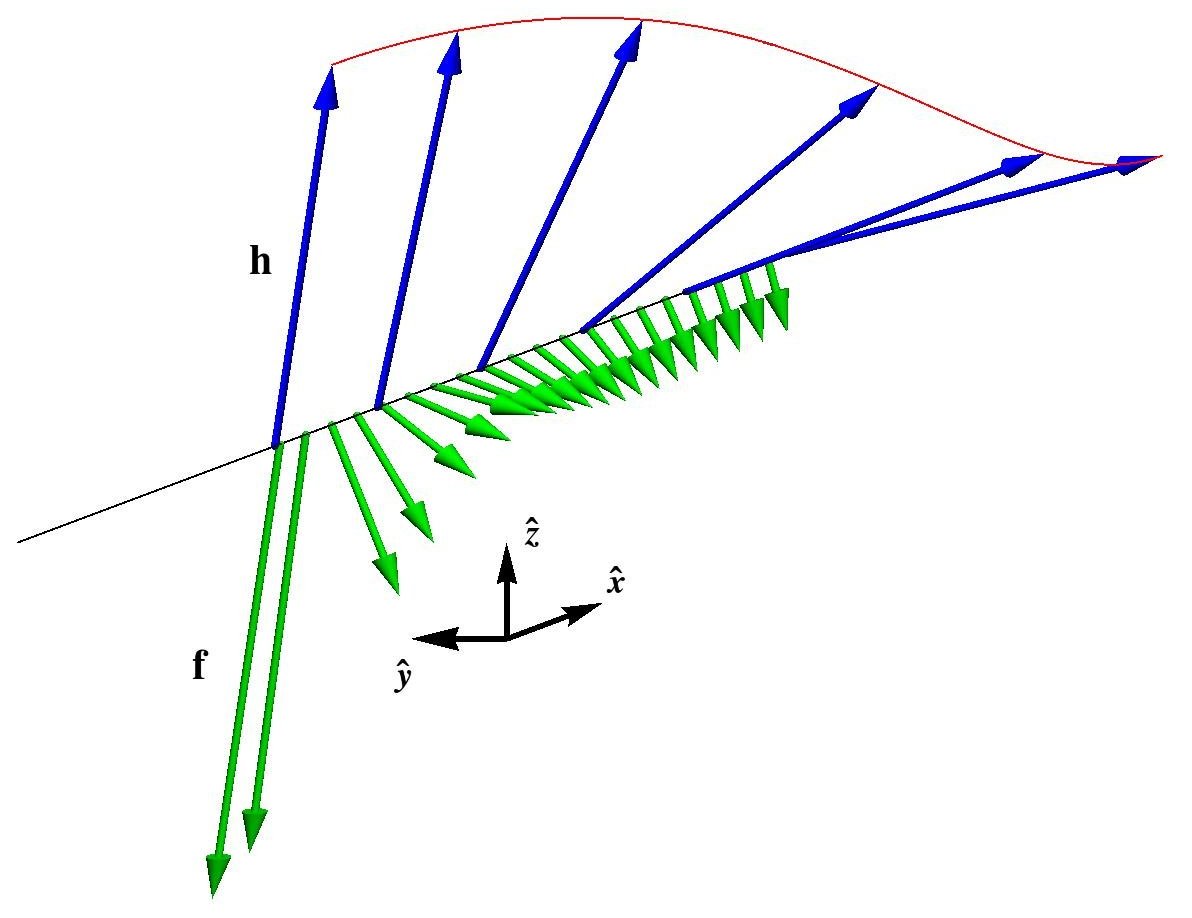}
\end{center}
\caption[Domain wall magnetization and Gor'kov vector in the XS]{\label{bakerFig2} (color online). Calculated domain wall magnetization profile $\mathbf{h}(x)$ (blue) and Gor'kov vector ${\bf f}(x)$ (green) in the XS. The S/XS (XS/S) interface is on the very left (right) of the figure. The vector $\mathbf{f}$ is obtained for $\omega_0$ and has been scaled for clarity using $\mathbf{f}_N =\eta \ln\left|M\right|^2 \,\left(x,\sin\gamma,-\cos\gamma\right)$ where $\gamma$ is the angle of the Gor'kov function with respect to $-\mathbf{\hat z}$, and $\eta$ is an arbitrary scaling parameter set to $\eta=0.1$. $t_{\rm Co}/\delta_{\rm Co} = 2.5$, $t_{\rm Py}/\delta_{\rm Py} = 0.1$.}
\end{figure}
Figure \ref{bakerFig2} shows $\mathbf{f}(x)$ obtained from Eqs.~(\ref{usadeltheta},\ref{usadelM}) starting from the S/XS interface (left side of the figure) and the magnetization vector $\mathbf{h}$ for an example of a partial domain wall. A similar figure is obtained starting at the XS/S interface and both results can be added in the wide limit considered here.
 Due to the FFLO effect, the Gor'kov vector function (green arrows with variable length) is anti-aligned with the magnetization at the interface (blue arrows assumed with constant magnitude). This indicates that only $S_z = 0$ singlet and triplet components are present. Deeper into the XS, the Gor'kov vector rotates and decreases in magnitude due to the rotation of the magnetization $\mathbf{h}(x)$. For this particular tuning of the domain wall only the component of $\mathbf{f}$ perpendicular to ${\bf h}$ is left near the inflection point of $\phi(x)$ close to the center of the domain wall. The latter indicates the presence of the LRTC ($S_z = \pm 1$) \cite{ivanovPRB06}.  There remains a small component of ${\bf f}$ parallel to ${\bf h}$ that is associated with the short-range triplet and singlet contributions ($S_z = 0$) that will be discussed later \cite{eschrigPT11}.

The physical understanding of the generic behavior of the order parameter $\mathbf{f}$ in the XS domain wall shown in Fig.~\ref{bakerFig2} is that as singlet Cooper pairs penetrate into the F the $S_z = 0$ triplet component is generated first and has a maximum close to the interface. Then, the rotation of the magnetization partially transforms the $S_z = 0$ triplet component into an LRTC with $S_z \neq 0$. From the figure and the calculations we conclude that to observe the largest possible LRTC the magnetization profile near the interface must be tuned to generate a maximal $S_z = 0$ triplet component followed by a region where the twist of the domain wall transforms the $S_z = 0$ component into an $S_z\neq 0$ LRTC. This confirms statements made in Refs. \cite{ivanovPRB06,buzdinPRB11}. We emphasize that the required magnetization profile for the observation of an $S_z\neq 0$ LRTC is highly non-linear with fairly constant values over the characteristic length $\xi_F$ near the interfaces that rotates rapidly thereafter.  This is the case of the XS, and contrasts with the stiff functional form $h(x) = \cos(Qx)$ usually introduced to model a domain wall \cite{bergeretPRL01,champelPRB05,alidoustPRB10,alidoustPRB10b}.

%%%%%%%%%%%%%%%%%%%%%%%%%%%%%%%%%%%%%%%%%%%
%
\begin{figure}[htb] 
\begin{center}
\includegraphics[scale=0.15]{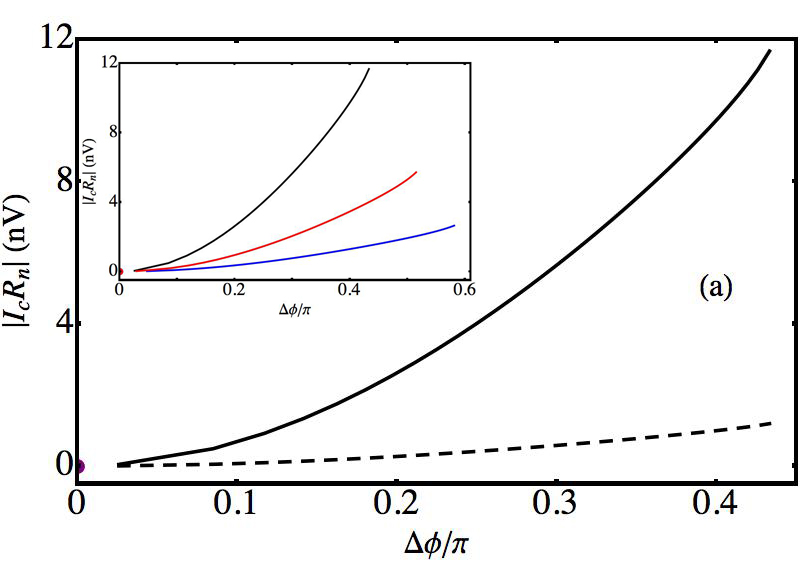}
\includegraphics[scale=0.15]{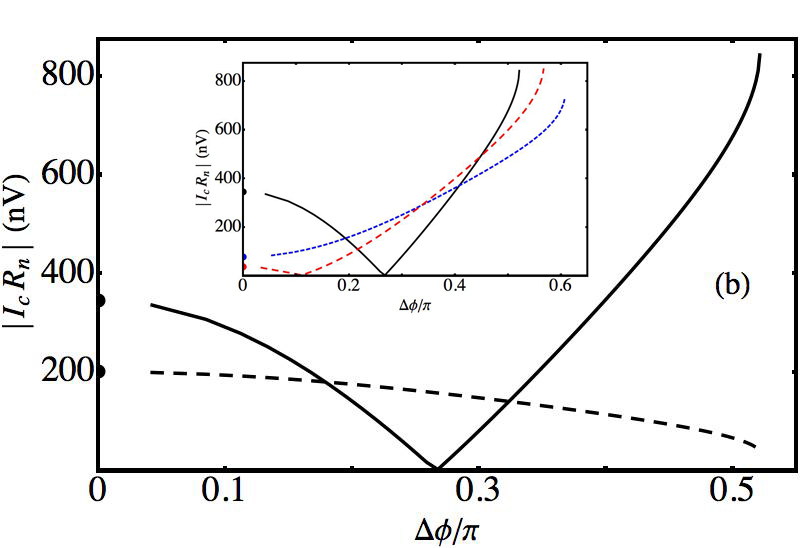}
\end{center}
\caption[Josephson current as a function of domain wall twist]{\label{bakerFig3} (color online). Josephson critical current (expressed as a potential $|V_c| = |I_c R_N|$) as a function of the domain wall twist in the XS $\Delta\phi / \pi$ (see text) for (a) Co/Py and (b) Ni$_3$Mn/Ni. The black dashed lines are the contribution of the $S_z = 0$ component whereas the solid lines represent the total current. The appearance of a current in (a) and its change of sign in (b) are due to the emergence of the LRTC with the coiling of the domain wall. Inserts: Total current for various thicknesses of the XS: (a) $t_{\rm Co}/\delta_{\rm Co} = 2$, 2.5, and 3 (top to bottom) and (b)  $t_{\rm Ni_3Mn}/\delta_{\rm Ni_3Mn} = 2$ (solid black), 2.25 (dashed red), 2.5 (dotted blue). Parameters are those of Nb and $t_{\rm Py}/\delta_{\rm Py} = 0.1$, $K_{\rm CoPy}/K_{\rm Py} = 625$, $h_\mathrm{Co}/\pi T_c=14$, $h_\mathrm{Py}/\pi T_c=8$, $t_{\rm Ni}/\delta_{\rm Ni} = 0.1$, $K_{\mathrm{Ni}_3\mathrm{Mn}}/K_\mathrm{Ni} = 10^3$, $h_{\mathrm{Ni}_3\mathrm{Mn}}/\pi T_c = 4.5$, and $h_\mathrm{Ni}/\pi T_c = 4$.}
\end{figure}
The precise knowledge of the Green and Gor'kov functions determined from the known analytic form of the XS allows the calculation of the Josephson critical current through the wide S/XS/S junction using Eq.~\eqref{IcRN}. This leads to the first prediction shown in Fig.~\ref{bakerFig3} where the Josephson voltage $V_c = I_c R_N$ ($R_N$ is the normal state resistance of the junction) is plotted against the normalized twist angle of the domain wall across the XS $\Delta\phi/\pi \equiv \left[\phi(-t_1) - \phi(t_2)\right]/\pi$.

Fig.~\ref{bakerFig3}a is obtained for the strong XS Co/Py with a thickness much larger than $\xi_F$ ($d_F/\xi_F =  22.5, 25$, and $27.5$, $d_F = t_1 + t_2$). Hence, in absence of a domain wall when $\Delta\phi/\pi = 0$, there is no measurable current flowing through the homogeneous magnetic Josephson junction. This reproduces the expected behavior of an S/F/S junction in the wide F limit. The remarkable effect occurs as one induces the domain wall into the XS ($0< \Delta\phi/\pi \leq 1$): A current appears and increases with the winding of the domain wall! From the discussion above, the growing current with increasing $\Delta\phi$ has its origin in the emergence of a LRTC.  This is substantiated by the fact that the singlet component to the current also plotted on the figure is an order of magnitude smaller.  It is noteworthy that an $S_z = 0$ component of the current is observed at all in such wide junction since this is the so-called short range component expected to decay over distances determined by $\xi_F$ which is of the order of the nanometer for Fig.~\ref{bakerFig3}a. The existence of this contribution is related to the continuous rotation of the quantization axis  and the mixing of the individual spin components.

The main implication of Fig.~\ref{bakerFig3}a is that an increasing current with the winding of the domain wall offers a new way to prove the existence of an LRTC generated by the inhomogeneous magnetization in the XS. Finally, the figure inset also shows that the current decreases with increasing thickness of either parts of the XS. This is due to the associated decrease in curvature of $\mathbf{h}(x)$ [or $\phi(x)$] of the domain wall with increasing thickness of the XS, and the resulting damping of the $S_z = 0$ triplet component near the S/XS interfaces.

%%%%%%%%%%%%%%%%%%%%%%%%%%%%%%%%%%%%%%%%%%%
The second prediction resulting from the study of the system with a tunable domain wall is shown in Fig.~\ref{bakerFig3}b. We plot $|V_c|$  as a function of domain wall twist for the weak XS Ni$_3$Mn/Ni. We note three salient features of this result. The total current (black solid line) undergoes a $0-\pi$ transition as shown by the $V-$shape curve reaching zero at $\Delta\phi/\pi \approx 0.26$. This current reversal is qualitatively different from those discussed earlier in the literature in that the present transition is {\it solely due to the emergence of the LRTC.} The voltage resulting from the singlet component of the order parameter here never changes sign (monotonically decreasing black dashed line). The reversal of the current with increasing twist of the domain wall results from the growing $S_z = \pm 1$ component current opposing the $S_z=0$ contribution.  Note that the presence of a current for $\Delta \phi = 0$ is due to the fact that the Ni$_3$Mn/Ni is a weak XS, revealing the importance of both the hardness and the magnitude of the magnetization of the Fs composing the XS.

The second feature displayed in the inset of Fig.~\ref{bakerFig3}b is the shift of the $0-\pi$ transition to smaller torsions of the magnetization profile with increasing thickness of the XS. This shift emphasizes that the triplet $0-\pi$ transition cannot be observed for arbitrary thicknesses of the XS.  For a thick enough soft F layer, the $0-\pi$ transition disappears and we recover the behavior of Fig.~\ref{bakerFig3}a. 

The last feature we point out in Fig.~\ref{bakerFig3}b is particularly interesting for experimental endeavors: The $0 - \pi$ transition can be observed in {\it one and the same sample} by tuning the domain wall in the XS.  This contrasts with all previous models and experiments in which the $0-\pi$ transition was predicted and observed as a function of the variable thickness of the homogeneous F in the S/F/S junction and therefore required a different sample for each thickness.

\begin{figure}[htb]
\begin{center}
\includegraphics[scale=0.19]{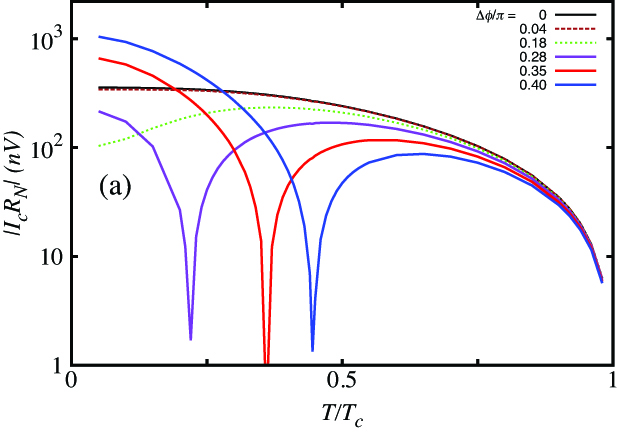}
\includegraphics[scale=0.19]{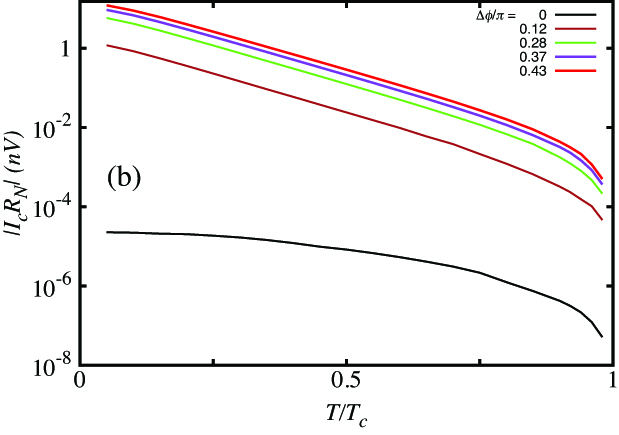}
\end{center}
\caption[$0-\pi$ transition with temperature]{\label{bakerFig4} (color online). Temperature dependence of the critical current $|I_cR_N|$ for different twists of the magnetization.  (a) Current for the weak XS Ni$_3$Mn/Ni with $t_{\rm Ni_3Mn}/\delta_{\rm Ni_3Mn} = 2$.  One observes a $0-\pi$ transition for large enough twisting of the domain wall. (b) Current for the strong XS Co/Py with $t_{\rm Co}/\delta_{\rm Co} = 2$. No $0-\pi$ transition is observed in this case. Note the logarithmic scale of the ordinates. Parameters as in Fig.~\ref{bakerFig1}.}
\end{figure}

%%%%%%%%%%%%%%%%%%%%%%%%%%%%%%%%%%%%%%%%%%%
In Fig.~\ref{bakerFig4}, we analyze the temperature dependence of the Josephson current for representative partial domain walls in XSs with weak (Fig.~\ref{bakerFig4}a) and strong (Fig.~\ref{bakerFig4}b) ferromagnetism and using the BCS temperature dependence of the pairing potential $\Delta(T) = \Delta_{\rm BCS} \sqrt{1-T^2/T_c^2}$.
The third prediction is that the {\it triplet} $0-\pi$ transition seen in Fig.~\ref{bakerFig3}b can also be induced with temperature. Fig.~\ref{bakerFig4}a shows that in absence of a domain wall the voltage decreases monotonically with temperature up to the critical temperature $T_c$ where it vanishes. As one induces the domain wall, a $0-\pi$ transition appears at low temperature. This transition moves to higher temperatures with increasing twist of the magnetic profile, reflecting a stronger triplet component of the order parameter.

Contrasting with this result, the strong XS Co/Py displays no $0-\pi$ transition (Fig.~\ref{bakerFig4}b) because the singlet component is very small (Fig.~\ref{bakerFig3}a). This is another consequence of different anisotropy ratios and magnitudes of the magnetization in the XS. Fig.~\ref{bakerFig4}b reveals three further interesting properties. First, the magnitude of the critical current decays exponentially fast over a large temperature range (linear decay on the figure). Second, the higher the twist, the higher the voltage and the more linear the curve on the log-linear scale. Finally, in accordance with the wide limit considered here the critical temperature remains unchanged \cite{degennesParks69}.

%%%%%%%%%%%%%%%%%%%%%%%%%%%%%%%%%%%%%%%%%%%
Figs.~\ref{bakerFig3} and \ref{bakerFig4} emphasize that the triplet $0-\pi$ transition is not observable in just any S/XS/S junction.  The intrinsic properties of the XS and in particular the anisotropy ratio, the thicknesses of the hard and soft Fs, and the magnitude of the magnetization are determinant parameters.  A large anisotropy ratio and thicknesses chosen to optimize the $S_z = 0$ triplet formation near the S/XS interfaces offer favorable conditions to observe the above phenomena.

Three corollaries of our analysis are worth mentioning. The LRTC can actually be observed in an FF' bilayer system provided that the two Fs interact magnetically, as is the case of the XS. Our proposed bilayer allows flexible modeling of different FF' systems realized experimentally since a tunable magnetization slip can be generated at the interface of the XS by placing a non-magnetic layer between the Fs to adjust their mutual magnetic interaction \cite{parkinPRL91}. The second corollary is that the triplet component can emerge in the XS with a domain wall stretching over the thickness of wide Josephson junctions that far exceed the coherence length $\xi_F$.  Finally, the domain wall generated in the XS is not symmetric. Our calculations  demonstrate that it is not necessary to have a symmetric multilayer to generate the LRTC. The main requirement is that the Gor'kov functions of both Ss overlap in the magnetic barrier.

The multilayer structure involving an XS proposed here has motivated the work done in Ref.~\cite{guPRB10} and also studied in Ref.~\cite{zhuPRL13} although these papers focused on the variation of $T_c$.  The study of the Josephson current through the XS is an experimental challenge that remains open.

%%%%%%%%%%%%%%%%%%%%%%%%%%%%%%%%%%%%%%%%%%%
In conclusion, the magnetic Josephson junction comprised of two singlet-pairing Ss and a magnetic XS is unique in that it allows the tuning of the long range triplet supercurrent through wide junctions by tuning the domain wall in the XS. Using an exact analytic expression for the domain wall in the XS \cite{billJMMM04} we offer insight in the relevant parameters and inhomogeneity required for the emergence of the LRTC. We propose an experiment in which the absence of domain wall inhibits the Josephson current through a wide junction, whereas a triplet current emerges with the increasing twist of the domain wall. We also predict that in junctions involving XSs with weak magnetization ({\it e.g.} Ni$_3$Mn/Ni) the tuning of the domain wall allows for the observation of a {\it triplet} $0-\pi$ transition in one and the same sample with fixed thicknesses of the magnetic layers. A similar transition of the triplet current is also observable at fixed twist of the magnetization by varying the temperature.

\vspace*{0.5cm}
%%%%%%%%%%%%%%%%%%%%%%%%%%%%%%%%%%%%%%%%%%%
We gratefully acknowledge funding provided by the National Science Foundation (DMR-0907242) and the Army Research Laboratory.  Early stages of the work were also funded by the Research Corporation. T.E.B.~gratefully acknowledges the Graduate Research Fellowship at CSU Long Beach.

%%%%%%%%%%%%%%%%%%%%%%%%%%%%%%%%%%%%%%%%%%%

\end{document}